\documentclass[aps,letterpaper,showpacs]{revtex4}
\usepackage{graphicx}
\begin{document}

\title{Mean field description of the Dicke Model}






\author{Jorge G. Hirsch,  Octavio Casta\~nos, Ram\'on L\'opez-Pe\~na, and Eduardo Nahmad-Achar}

\affiliation{Instituto de Ciencias Nucleares,
Universidad Nacional Aut\'onoma de M\'exico \\
Apdo. Postal 70-543, Mexico D. F., C.P. 04510}

\begin{abstract}
A mean field description of the Dicke model is presented, employing the Holstein-Primakoff realization of the angular  momentum algebra. It is shown that, in the thermodynamic limit, when the number of atoms interacting with the photons goes to infinity the energy surface takes a simple form, allowing for a direct description of many observables.
\end{abstract}

\maketitle

\section{Introduction}

The Dicke Hamiltonian has been successfully employed to describe the superradiance when 
two-level atoms interact with a one mode electromagnetic field  \cite{dicke}, in superconducting circuit QED systems \cite{Nat10}, and in the open experimental setup  \cite{baumann} where the phase transitions in a superfluid gas inside an optical cavity and circuit QED was observed \cite{Dim07, nagy}. The Holstein-Primakoff realization of the quasi-spin operators representing the collective atomic excitations has been widely employed in its description \cite{Ema03,Dim07,nagy,Nat10}. 

The expectation values of the energy per particle, the number of photons per particle and the fraction of excited atoms, has been calculated employing a truncated version of the Dicke Hamiltonian in the normal and superradiant  phases in the thermodynamic limit, and found to be in good agreement with the numerical estimations of these values even for  a small number of atoms \cite{Ema03}. In this contribution we show that these expressions can be found directly from a mean field description of the Dicke model. They also reproduce the expressions found employing SU(2) coherent states, but fail in describing the fluctuations in the number of photons and atoms, which requires the use of symmetry-adapted coherent states \cite{Cas10,Cas11}.

 The Dicke Hamiltonian describes the collective interaction of $N$ two-level atoms with energy separation equal to $\hbar \tilde\omega_A$ with a one mode radiation field of frequency $\tilde\omega_F$. 
It is given by
	\begin{equation}
		H_{D}= a^{\dagger}a + \omega_{A}\,J_{z}
		+\frac{\gamma}{\sqrt{N}}\left(a^{\dagger} + a \right) \left( \,J_{-}
		+ \,J_{+}\right)\ ,
		\label{D}
	\end{equation}
where $\omega_A= \tilde{\omega}_A/\tilde{\omega}_F \geq 0$ is given in units of the frequency of the field, and $\gamma= \tilde{\gamma}/\tilde{\omega}_F$ is the (adimensional) coupling parameter. The operators $a$, $a^{\dagger}$, denote the one-mode annihilation and creation photon operators, respectively, $J_{z}$ the atomic relative population operator, and $J_{\pm}$ the atomic transition operators.

\section{Mean Field description}

In the literature it is customary to employ the Holstein-Primakoff representation of the angular momentum operators \cite{Kap91}, with $j=N/2$,
\begin{equation}
J_{+} = \, b^{\dagger} \sqrt{2j -  b^{\dagger}  b}, \,\,
J_{-} =  \sqrt{2j -  b^{\dagger}  b} \,\ b, \,\,
J_z = \, b^{\dagger}  b - j, 
\label{j}
\end{equation}
where new Bose operators  $b^{\dagger},  b$ are introduced, which obey the commutation relation 
$$
[b, b^{\dagger}] = 1,
$$
and the vacuum of the new bosons which satisfies $b |0_b\rangle = 0$ is $|0_b\rangle = |j, -j\rangle$.
Making these substitutions into $H_D$, Eq. (\ref{D}), the two-mode bosonic Hamiltonian becomes
\begin{eqnarray}
	H_{HP}=& \,a^{\dagger}a + \omega_{A}\, (b^{\dagger}  b - j)  \label{HHP}
	\label{hp}	\\ \nonumber
		&+ \gamma \left(a^{\dagger} + a \right) \left( \,b^{\dagger} \sqrt{1 - \frac{ b^{\dagger}  b}{2j}}
		+ \,\sqrt{1 - \frac{ b^{\dagger}  b}{2j}}\,\, b\right). 
\end{eqnarray}

The mean field  description of this Hamiltonian is easily obtained employing as a trial state the direct product of two
Heinseberg-Weyl coherent states \cite{Glau} $\vert \alpha \rangle$ and $\vert \beta \rangle$, which are eigenstates of the two bosonic annihilation operators

\begin{equation}
a\, \vert \alpha \rangle = \alpha \, \vert \alpha \rangle , \, \,
b \,\vert \beta \rangle = \beta \, \vert \beta \rangle ,
\label{ab}
\end{equation}
where $\alpha$ and $\beta$ are complex numbers, parameterized as $\alpha = \rho_a\, e^{i \phi_a}$, $\beta = \rho_b\, e^{i \phi_b}$, with $\rho_a, \rho_b \geq 0$ and $0 \leq \phi_a, \phi_b < 2 \pi$.

The explicit form of these coherent states is
\begin{equation}
|\zeta \rangle = e^{-\frac {|\zeta|^{2}}{2}}\sum^{\infty}_{n=0}\frac{(\zeta \, a^\dagger)^{n}}{n!} |0\rangle
=  e^{-\frac {|\zeta|^{2}}{2}}\sum^{\infty}_{n=0} \frac{\zeta^n} {\sqrt{n!}} |n \rangle,
\hbox{~~~~with~~~} \zeta = \alpha,\beta.
\label{coh}
\end{equation}    
In the above expression $|n\rangle$ is the normalized Fock state with $n$ bosons. 

To obtain the energy surface 
${\cal E}(\omega_A, \gamma, j ; \alpha, \beta) \equiv \langle \alpha \beta \vert H_{HP} \vert \alpha \beta \rangle $
some matrix elements are directly obtained from Eq.  (\ref{ab}). They are
\begin{equation}
\begin{array}{rl}
\langle \alpha | a^{\dagger} a | \alpha \rangle = | \alpha |^2 = \rho_a^2, &
 \langle \beta | b^{\dagger} b | \beta \rangle = | \beta |^2 = \rho_b^2, \\
 \langle \alpha | a^{\dagger} + a | \alpha \rangle = \alpha + \alpha^* = & 2 \, \rho_a \, cos{\phi_a}.
 \end{array}
 \end{equation}

More interesting is the evaluation of the matrix element of the last term in $H_{HP}$, which requires the use of the explicit form of the coherent state, Eq. (\ref{coh}). For the square root to make sense in the last term of Eq. (\ref{hp}), the maximum number of $b$ bosons must be limited to $2j$. 

\begin{equation}
\begin{array}{c}
\langle \beta | \,b^{\dagger} \sqrt{1 - \frac{ b^{\dagger}  b}{2j}}	+ \,\sqrt{1 - \frac{ b^{\dagger}  b}{2j}}\,\, b |\beta \rangle =\\
  e^{-|\beta|^{2}}  (\beta + \beta^*) \sum\limits^{2j}_{n=0}  \frac{|\beta|^{2n}  } {n!}  \sqrt{1 - \frac{n}{2j}} =\\
2 \, e^{-\rho_b^{2}} \, \rho_b \, cos{\phi_b} \sum\limits^{2j}_{n=0}  \frac{\rho_b^{2n}  } {n!}  \sqrt{1 - \frac{n}{2j}}   \end{array}
\end{equation}

Putting together the previous expressions, the final form of the energy surface is
\begin{equation}
{\cal E}(\omega_A, \gamma, j ; \alpha, \beta) = \rho_a^2 + ( \rho_b^2 -j )\, \omega_A +  4  \, \gamma \, \rho_a \rho_b \, \cos \phi_a \cos \phi_b  \, e^{-\rho_b^{2}} \, \sum\limits^{2j}_{n=0}  \frac{\rho_b^{2n}  } {n!}  \sqrt{1 - \frac{n}{2j}} .
\end{equation}

\section{The Thermodynamic Limit}

From the Holstein-Primakoff representation, Eq. (\ref{j}), it follows directly the Heisenberg-Weyl contraction of the SU(2) group. This contraction corresponds to the thermodynamic limit  $j \rightarrow \infty$, and in this case the SU(2) and Heisenberg-Weyl coherent states become equivalents \cite{Gilm}.

The function 
\begin{equation}
F(\rho,j) \equiv \langle \beta | \sqrt{1 - \frac{ b^{\dagger}  b}{2j}} |\beta \rangle = e^{-\rho_b^{2}} \, \sum\limits^{2j}_{n=0}  \frac{\rho_b^{2n}  } {n!}  \sqrt{1 - \frac{n}{2j}},
\end{equation}
has the asymptotic limit
\begin{equation}
\hbox{Lim}_{j \rightarrow \infty} F(\rho,j) = \sqrt{1 - {\frac{\rho_b^2}{2 j}}}.
\end{equation}
This expression can be ``guessed" doing the substitution
\begin{equation}
 \langle \beta |\sqrt{1 - \frac{ b^{\dagger}  b}{2j}} |\beta \rangle \rightarrow \sqrt{1 - \frac{ \langle \beta | b^{\dagger}  b  |\beta \rangle}{2j}} ,
 \end{equation}
which is equivalent to replace  $ \langle \beta | (b^{\dagger} b)^n | \beta \rangle$ by $ \langle \beta | b^{\dagger} b|\beta \rangle ^n$ in the series expansion of the square root. All the terms which are not in normal order are neglected in doing so. 

\begin{figure}[h]
\scalebox{1.5}{\includegraphics{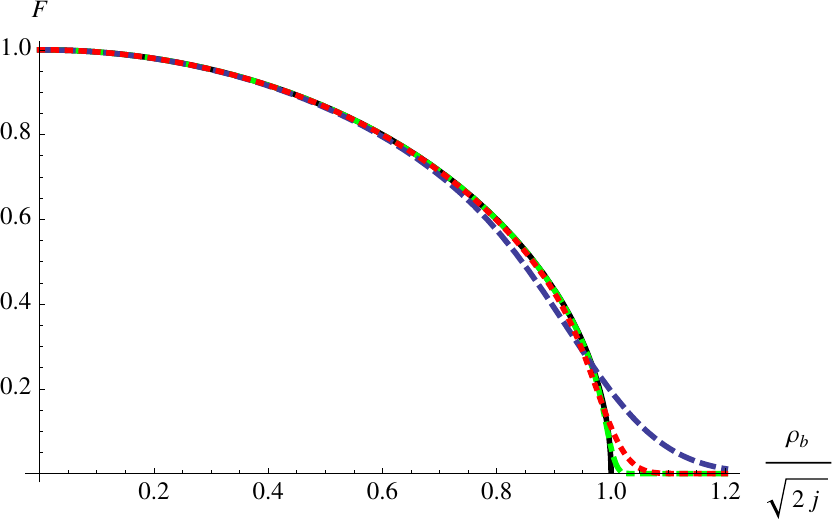}}
\qquad\qquad
\caption{\label{F}
(Color online) The function $F(\rho,j)$ evaluated as function of $\frac {\rho_b}{\sqrt{2j}}$, for j=10 (dashed blue line), 100 (dotted red line) and 1000 (green dot-dashed line), and its limit for  $j \rightarrow \infty$ (continuous black line) . }
\end{figure}

As it can be seen in Fig. \ref{F}, where $F(\rho,j)$ is evaluated as a function of $\frac {\rho_b}{\sqrt{2j}}$ and compared with the simpler limiting expression, the convergence is fast, and even for $j=100$ it is hard to distinguish between the different calculations.

In the thermodynamic limit, when the number of atoms goes to infinity, the energy surface ${\cal E}$ takes the simple form
\begin{equation}
{\cal E} = \rho_a^2 + ( \rho_b^2 -j )\, \omega_A +  4  \, \gamma \, \rho_a \rho_b \,\sqrt{1 - {\frac{\rho_b^2}{2 j}}} \cos \phi_a \cos \phi_b 
\label{et}
\end{equation}
		
For a given set of Hamiltonian parameters  $ \omega_{A}, \gamma$, the values $ \rho_{a}, \phi_{ac}, \rho_{b}, \phi_{b}$ which minimize this expression provide the mean field wave function, the best approximation to the exact ground state as a product wave function. They are obtained by solving the equations for the energy surface critical points
\begin{eqnarray}
\frac {\partial {\cal E}}{\partial \rho_{i}} = 0, ~~~
 \frac {\partial {\cal E}}{\partial \phi_{i}} = 0, ~~~~~~i= a,b.
 \end{eqnarray}
The explicit form of these equations is
\begin{eqnarray}
 \rho_{a}  +  2 \gamma \rho_{b} \sqrt{1 -  \frac{\rho_{b}^2}{2 j}} \cos \phi_{a} \cos \phi_{b} = 0 , \\ 
\gamma  \rho_{a}  \rho_{b}   \sqrt{1 -  \frac{\rho_{b}^2}{2 j}} \sin \phi_{a} \cos \phi_{b} = 0 ,\\
 \gamma  \rho_{a}\cos \phi_{a} \cos \phi_{b}  - \omega_A  \rho_{b}   \sqrt{1 -  \frac{\rho_{b}^2}{2 j}} = 0,  \\
\gamma  \rho_{a}  \rho_{b}   \sqrt{1 -  \frac{\rho_{b}^2}{2 j}} \cos \phi_{a} \sin \phi_{b} = 0 .  \end{eqnarray}
And the solutions associated with the minima of the energy surface are
\begin{eqnarray}
\left\{ 
\begin{array}{cc}
 \rho_{a} = 0,  &\phi_{a} \hbox{\small ~~undetermined,}   \\ \rho_{b} = 0,  &\phi_{b} \hbox{\small ~~undetermined,}  \end{array} \right.
 &\hbox{ if }  \gamma  < \gamma _c,  \\
\left\{ 
\begin{array}{cc}
 \rho_{a} = \gamma \sqrt{2 j }    \sqrt{1-\left(\frac{\gamma _c}{\gamma }\right)^4}, &\phi_{a} = 0, \\
\rho_{b} = \sqrt{j} \sqrt{1-\left(\frac{\gamma _c}{\gamma}\right)^2}, &\phi_{b} = \pi, \end{array} \right.&\hbox{ if }  \gamma  \ge  \gamma _c,
\end{eqnarray}
with
  $$
  \gamma _c = \frac{\sqrt{\omega_A}}{2}.
 $$
 Substituting these expressions in Eqs. (\ref{ab}) and (\ref{et}), the expectations values of the energy per particle, the number of photons per particle and the fraction of excited atoms are obtained, in full agreement with previous works \cite{Ema03,Cas10,Cas11}.

\section{Conclusions}

We have shown that a simple expression can be found for the energy surface of the Dicke Hamiltonian, employing a Holstein-Primakoff realization of the rotor algebra for the atomic sector, and using as trial states the boson coherent states associated with the photons and the atoms.  A variational treatment of the energy surface allows to obtain analytical expressions for some intensive observables, which reproduce those obtained previously with truncated versions of the Dicke Hamiltonian \cite{Ema03}. They also reproduce the expressions found employing SU(2) coherent states, but fail in describing the fluctuations in the number of photons and atoms, which requires the use of symmetry-adapted coherent states \cite{Cas10,Cas11}.

This work was partially supported by CONACyT-México, DGAPA- UNAM (projects IN102811 and 102109), and FONCICyT (project 94142).

\end{document}